\documentclass[12pt,preprint]{aastex}

\slugcomment{This manuscript is accepted by \textbf{Ap\&SS}}

\begin{document}

\title{Artificial viscosity in simulation of shock waves by smoothed particle hydrodynamics}

\author{M. Nejad-Asghar}

\affil{Department of Physics, Damghan University of Basic Sciences,
Damghan, Iran} \email{nasghar@dubs.ac.ir}

\author{A.R. Khesali, J. Soltani}

\affil{Department of Physics, Mazandran University, Babolsar, Iran}

\begin{abstract}
The artificial viscosity is reconsidered in smoothed particle
hydrodynamics to prevent inter-particle penetration, unwanted
heating, and unphysical solutions. The coefficients in the
Monaghan's standard artificial viscosity are considered as time
variable, and a restriction on them is proposed such that
avoiding the undesired effects in the subsonic regions. The shock
formation in adiabatic and isothermal cases are used to study the
ability of this modified artificial viscosity recipe. The computer
experiments show that the proposal appears to work and the
accuracy of this restriction is acceptable.
\end{abstract}


\section{Introduction}
Gas dynamical processes are believed to play an important role in
the evolution of astrophysical systems on all length scales. A
ubiquitous process in astrophysical fluid dynamics is the
behavior of gases subjected by shock waves. These are almost the
normal, rather than exceptional, type of astrophysical fluid
flows that happen very often in cases of astrophysical interest.
A number of examples are the onset of spiral structures and shock
fronts in accretion disks around compact objects (Lanzafame et
al.~2006), molecular cloud formation in shock-compressed layers
(Koyama and Inutsuka~2000), the rapid and strong increase in
pressure that can be produced by interaction of the fast wind and
the slow envelope of a planetary nebulae (Gurzadyan~1997), and so
on.

A powerful gridless particle method that was invented to solve
complex fluid-dynamical problems in astrophysics is smoothed
particle hydrodynamics~(SPH) (Lucy~1977, Gingold and
Monaghan~1977). The SPH method has a number of attractive
features such as its low numerical diffusion in comparison to
grid based methods. Whilst the SPH originally developed for
compressible flows, it has been extended to deal with
free-surface incompressible fluids (e.g., Monaghan~1994, Ellero
et al.~2007). A recent worthy review of the SPH methodology and
its applications can be found in Monaghan~(2005).

An adequate scenario for SPH application is the unbound
astrophysical problems, especially the behavior of gases
subjected to compression (shock waves). The magnitude of the
viscous term dose not affect the net shock-jump conditions, thus,
many numerical schemes implicitly or explicitly incorporate the
trick of artificial viscosity~(AV) for halting the ever-growing
steepening tendency produced by nonlinear effects. The first use
of viscosity in SPH equations was by Lucy~(1977) who introduced
an AV to present a slow build-up of acoustic energy from
integration errors in an SPH simulation. The AV is designed to
allow shock phenomena to be simulated, or simply to stabilize a
numerical algorithm. Indeed two forms of AV are applied in SPH
equations, the bulk and the von Neumann-Richtmeyer viscosity,
respectively. They prevent inter-particle penetration, allow
shocks to form and damp the post-shock oscillations. Although
many different functional forms for the AV have been proposed
(see, e.g., Liu \& Liu~2003), all contain problem sensitive
parameters that are often set in a somewhat arbitrary manner.
Here we try firstly to provide a guide to the relevant
prescriptions for some AV that have been considered before.

A more effective viscosity which may conserves linear and angular
momentum was suggested by Monaghan and Gingold (1983). They
devised a viscosity by simple arguments about its form and its
relation to gas viscosity. The viscous term between two
particles, denoted by $\Pi_{ab}$, is added to the pressure term
of SPH equations. When two particles approach each other, the AV
produces a repulsive force between them. When they recede from
each other the force is attractive (Monaghan~1989).

An undesirable aspect of the Monaghan's standard AV is that it
can introduce considerable shear viscosity into the flow. This
problem may be reduced by a bulk viscosity in a general-purpose
code of Hernquist and Katz (1989) for evolving three-dimensional,
self-gravitating fluids in astrophysics. Clearly the artificial
viscous dissipation increases the Reynolds number of a flow,
artificially, with the result that, for example, the
Kelvin-Helmholtz shear instabilities are heavily diffused. In
this way a von Neumann stability analysis of the SPH equations
along with a critical discussion of various parts of the
algorithm was investigated by Balsara~(1995). He suggested
reducing viscous dissipation by multiplying $\Pi_{ab}$ by a
symmetric factor.

Another problem of viscosity is that although AV is approximately
successful for handling shocks but it can be too large in other
parts of the flow. For this purpose, a switch to reduce the AV
away from shocks was given by Morris and Monaghan (1997, hereafter
MM97). They introduce the idea of time-varying switch which fits
more naturally with a particle formulation. Each particle has a
viscosity parameter which evolves according to a simple source
and decay equation. The source causes the switch to grow when the
particle enters a shock and the decay term causes it to decay to
a small value beyond the shock.

In the present study we combine the Monaghan's standard AV with
the time-varying coefficients like that of MM97's switch. we
propose a modification to the time-dependent AV prescription
designed by MM97 so that the aim is to maintain the
shock-simulating capabilities of SPH. For this purpose we
optimize the source term so that the AV is restricted to
supersonic velocities in regions that are under compression.
Section~2 is devoted to the SPH method and the suggested changes
in the AV. These modifications have been tested in \S3 for the
one dimensional adiabatic and isothermal shock problems. Finally,
a summary with conclusion is given in section~4.

\section{The SPH methodology}
The SPH was invented to simulate nonaxisymmetric phenomena in
astrophysics~(Lucy 1977, Gingold \& Monaghan 1977). In this
method, fluid is represented by $N$ discrete but
extended/smoothed particles (i.e. Lagrangian sample points). The
particles are overlapping, so that all the involved physical
quantities can be treated as continuous functions both in space
and time. Overlapping is represented by the kernel function,
$W_{ab} \equiv W(\textbf{r}_a-\textbf{r}_b,h_{ab})$, where
$h_{ab} \equiv (h_a+h_b)/2$ is the mean smoothing length of two
particles $a$ and $b$. The continuum equations are~(Monaghan 1992)
\begin{equation}
\rho_a=\sum_b m_b W_{ab}
\end{equation}
\begin{equation}
\frac{d\textbf{v}_a}{dt}=-\sum_b m_b (\frac{p_a}{\rho_a^2}+
\frac{p_b}{\rho_b^2}+ \Pi_{ab}) \nabla_a W_{ab}
\end{equation}
\begin{equation}
\frac{du_a}{dt}=\frac{1}{2} \sum_b m_b (\frac{p_a}{\rho_a^2}+
\frac{p_b}{\rho_b^2}+ \Pi_{ab}) \textbf{v}_{ab} \cdot \nabla_a
W_{ab}
\end{equation}
where $\textbf{v}_{ab}\equiv \textbf{v}_a- \textbf{v}_b$, the AV
between particles $a$ and $b$ is presented by $\Pi_{ab}$, and the
other notations have their usual meanings.


\subsection{Viscosity}

In order to simulate the shock problems of hydrodynamics, special
treatments or methods are required to allow the algorithms to be
capable of modeling shock waves, or else the simulation will
develop unphysical oscillations in the numerical results around
the shocked regions. A shock wave is not a true physical
discontinuity, but a very narrow transition zone whose thickness
is usually in the order of a few molecular mean free paths.
Application of the conservation of mass, momentum, and energy
conditions across a shock front requires the simulation of
transformation of kinetic energy into heat energy. Physically
this energy transformation can be represented as a form of
viscous dissipation. This idea leads to the development of the
von Neumann-Richtmyer AV that is needed only to be present during
material compression,
\begin{equation}\label{quadvisc}
\Pi_1=\cases{
       a_1 (\triangle x)^2 \rho (\nabla . \mathbf{v})^2 &
       $\nabla . \mathbf{v} < 0$ \cr
       0  & $\nabla . \mathbf{v} \geq 0$,}
\end{equation}
where $a_1$ is an adjustable non-dimensional constant. It is found
that adding the following linear AV term,
\begin{equation}\label{linvisc}
\Pi_2=\cases{
       a_2 (\triangle x) c \rho (\nabla . \mathbf{v}) &
       $\nabla . \mathbf{v} < 0$ \cr
       0  & $\nabla . \mathbf{v} \geq 0$,}
\end{equation}
where $c$ is the sound speed and $a_2$ is another non-dimensional
constant (Liu and Liu~2003), has the advantage of further
smoothing the oscillations that are not totally dampened by the
quadratic AV term, equation (\ref{quadvisc}).

The most widely used AV in SPH is that of Monaghan (1989), which
not only provides the necessary dissipation to convert kinetic
energy into heat at the shock front, but also prevent unphysical
penetration for particles approaching each other. The detailed
formulation is as follows:
\begin{equation}\label{monag}
\Pi_{ab}=\cases{
       \frac{-\alpha v_{sig} \mu_{ab}
       +\beta \mu_{ab}^2}{\bar{\rho}_{ab}}, &
       if $\textbf{v}_{ab}.\textbf{r}_{ab}<0$,\cr
       0 , & otherwise,}
\end{equation}
where $\bar{\rho}_{ab}= \frac{1}{2}(\rho_a+\rho_b)$ is an average
density, $\alpha\sim 1$ and $\beta\sim 2$ are the artificial
coefficients, and $\mu_{ab}$ is defined as its usual form
\begin{equation}
\mu_{ab}=\frac{\textbf{v}_{ab} \cdot\textbf{r}_{ab}}{\bar{h}_{ab}}
\frac{1}{r_{ab}^2/\bar{h}_{ab}^2+\eta^2}
\end{equation}
with $\eta\sim 0.1$ and $\bar{h}_{ab}= \frac{1}{2}(h_a+h_b)$. The
signal velocity, $v_{sig}$, is
\begin{equation}
v_{sig}=\frac{1}{2}(c_a+c_b)
\end{equation}
where $c_a$ and $c_b$ are the sound speed of particles.

Since the Monaghan type AV introduces a shear viscosity into the
flows especially in regions away from the shock, an AV depending
on the divergence of the velocity field was employed by Hernquist
and Katz (1989),
\begin{equation}
\Pi_{ab}=\frac{q_a}{\rho_a^2}+\frac{q_b}{\rho_b^2},
\end{equation}
where
\begin{equation}
q_a=\cases{
       \alpha h_a \rho_a c_a |\nabla . \mathbf{v}|_a +
       \beta h_a^2 \rho_a |\nabla . \mathbf{v}|_a^2 &
       $\nabla . \mathbf{v}<0$ \cr
       0  & otherwise.}
\end{equation}
This AV can lead to spurious results if particles $a$ and $b$ are
receding; i.e., such terms will contribute viscous artificial
cooling. To avoid this problem, Hernquist and Katz (1989) set
$\Pi_{ab}=0$ if $\mathbf{v}_{ab} . \mathbf{r}_{ab} > 0$; a
decision which may, in some cases, add small amounts of shear
viscosity. In this approach, determining $\Pi_{ab}$ by referring
to binary interaction, effectively, gives an idea about the
molecular bulk fluid descriptions.

Other view on the Monaghan's standard AV was also proposed by
Balsara~(1995) who developed a modification of the AV term that
approximately sets the dissipation to zero in regions of pure
shear, while leaving it unaffected in regions of compression.
This is done by suggestion
\begin{equation}
\Pi_{ab}=\left( \frac{p_a}{\rho_a^2}+\frac{p_b}{\rho_b^2} \right)
(-\alpha \mu_{ab} + \beta \mu_{ab}^2),
\end{equation}
where a switch in the form of a multiplicative factor ($\mu_{ab}
\rightarrow \mu_{ab}f_{ab}/\bar{c}_{ab}$) was introduced. The factor
$f_{ab}$ is the average of $f_a$ and $f_b$, where
\begin{equation}
f_a= \frac{|\nabla . \mathbf{v}|_a}{|\nabla . \mathbf{v}|_a +
|\nabla \times \mathbf{v}|_a + 10^{-4} c_a/h_a}.
\end{equation}
The function $f_a$ acts as a switch, approaching unity in regions
of compression ($|\nabla . \mathbf{v}|_a >> |\nabla \times
\mathbf{v}|_a$) and vanishing in regions of large vorticity
($|\nabla \times \mathbf{v}|_a >> |\nabla . \mathbf{v}|_a$).
Consequently, this AV has the advantage that it is suppressed in
shear layers.

Meanwhile, MM97 carried the process further by introducing the
idea of time-varying coefficients which fits more naturally with
a particle formulation. In this method, the AV is given by
\begin{equation}\label{morris}
\Pi_{ab} = - \frac{K_{ab} v_{sig} \mathbf{v}_{ab}\cdot
\mathbf{r}_{ab}}{\bar{\rho}_{ab} |\mathbf{r}_{ab}|},
\end{equation}
where $K_{ab} = (K_a+K_b)/2$ is a variable switch which should
change with time according to the conditions the particle is in,
becoming large at shocks, but relaxing back to a small value when
the flow is calmer. Each particle $a$ has a viscosity switch
which evolves according to a simple source,
\begin{equation}\label{source}
\mathcal{S}_a=\max(-\nabla\cdot \mathbf{v}_a , 0),
\end{equation}
and decay equation,
\begin{equation}
\frac{dK_a}{dt} = - \frac{K_a - K_{\mathrm{min}}}{\tau_a} +
\mathcal{S}_a,
\end{equation}
such that in the absence of source term $\mathcal{S}_a$, the
switch $K_a$ decays to a value $K_{\mathrm{min}}$ over a time
scale $\tau_a \equiv h_a / (\mathcal{C} v_{sig})$. The
dimensionless parameter $\mathcal{C}$ has a value $0.1 <
\mathcal{C} < 0.2$. The source causes the switch to grow when the
particle enters a shock and the decay term causes it to decay to
a small value beyond the shock.

\subsection{Restricted AV}

In SPH the formation of shocks is mostly an effect of the von
Neumann-Richtmeyer viscosity. To avoid the undesirable effects in
the subsonic region, we propose to restrict the use of it to those
regions. This modification will allow shocks to form, and prevent
inter-particle penetration at supersonic velocities. When the gas
has reached equilibrium, the pressure force prevents further
compression. This process is not possible with source term
(\ref{source}) because of supersonic relative velocities of the
particles at different regions of the shock. The AV as outlined
by equation (\ref{morris}) will decelerate and heat the gas, and
prevent enough compression. To weaken this problem, we use the
following lemma.

Since $\rho_a \propto h_a^{-\nu}$ where $\nu$ is the dimension of
the problem, the time derivative for any particle can be written
as
\begin{equation}
\dot{\rho_a} = - \frac{\nu \rho_a \dot{h}_a}{h_a}.
\end{equation}
This relation can be used to decide whether a particle follows the
fluid or if the source term of AV is necessary. If two particles are
approaching each other at velocity exceeding the signal velocity,
that is if
\begin{equation}
r_{ab} ( \frac{\dot{\rho}_a}{\rho_a} - \frac{\nu \dot{h}_a}{h_a})
> v_{\mathrm{sig}},
\end{equation}
AV is necessary to prevent inter-particle penetration. We
therefore propose to restrict the source term of AV with defined
frequency as
\begin{equation}
\omega_a \equiv \frac{\dot{\rho}_a}{\rho_a} - \frac{\nu
\dot{h}_a}{h_a} - \frac{v_{\mathrm{sig}}}{r_{ab}}.
\end{equation}
A good way to do this for a particle is to use the restricted
source term $\mathcal{S}_a$, instead of equation (\ref{source}),
as follows
\begin{equation}
\mathcal{S}_a=\max (-\nabla . \mathbf{v}_a , \omega_a ,0).
\end{equation}

Base on a method similar MM97, we use $\alpha$ and $\beta$ in the
standard AV, equation (\ref{monag}), in form of variables with
respect to time,
\begin{equation}\label{alpha}
\frac{d \alpha_a}{dt} = - \frac{\alpha_a - \alpha_{\mathrm{min}}}
{\tau_a} + z_\alpha \mathcal{S}_a,
\end{equation}
and
\begin{equation}\label{beta}
\frac{d \beta_a}{dt} = - \frac{\beta_a - \beta_{\mathrm{min}}}
{\tau_a} + z_\beta \mathcal{S}_a,
\end{equation}
respectively, where the parameters $z_\alpha$ and $z_\beta$ are
chosen to regulate the effect of source term such that the
\textit{heat production} and \textit{post-shock oscillations} are
controlled in the numerical simulations.

\section{Shock formation test}

\subsection{Analytical formulation}

As outlined before, an extremely important problem is the behavior
of gases subjected to compression waves. This happens very often
in the cases of astrophysical interests. For example, a small
region of gas suddenly heated by the liberation of energy will
expand into its surroundings. The surroundings will be pushed and
compressed, thus, a shock front is formed. Conservation of mass,
momentum, and energy across a shock front is given by the
Rankine-Hugoniot conditions (Dyson and Williams 1997)
\begin{equation}\label{e:rh1}
\rho_1 v_1=\rho_2 v_2
\end{equation}
\begin{equation}
\rho_1 v_1^2+ K\rho_1^\gamma =\rho_2 v_2^2+ K\rho_2^\gamma
\end{equation}
\begin{equation}\label{e:rh3}
\frac{1}{2}v_1^2 + \frac{\gamma}{\gamma-1} K \rho_1^{\gamma-1}=
\frac{1}{2}v_2^2 + \frac{\gamma}{\gamma-1} K \rho_2^{\gamma-1} +Q
\end{equation}
where the equation of state, $p=K\rho^\gamma$, is used. In adiabatic
case, we have $Q=0$, and for isothermal shocks, we will set
$\gamma=1$.

We would interested to consider the collision of two gas sheets with
velocities $v_0$ in the rest frame of the laboratory. In this
reference frame, the post-shock will be at rest and the pre-shock
velocity is given by $v_1=v_0+v_2$, where $v_2$ is the shock front
velocity. Combining equations (\ref{e:rh1})-(\ref{e:rh3}), we have
\begin{equation}\label{e:v2}
v_2=a_0[-\frac{b}{2}+\sqrt{1+\frac{b^2}{4}+(\gamma-1)
(\frac{M_0^2}{2}-q)}]
\end{equation}
where $a_0\equiv \gamma K\rho_1 ^{\gamma-1}$ is the sound speed,
$M_0\equiv v_0/a_0$ is the Mach number, $b$ and $q$ are defined as
\begin{equation}
b\equiv \frac{3-\gamma}{2}M_0+ \frac{\gamma-1}{M_0}q,\quad q\equiv
\frac{Q}{a_0^2},
\end{equation}
respectively. Substituting (\ref{e:v2}) into equation
(\ref{e:rh1}), density of the post-shock is given by
\begin{equation}\label{e:den}
\rho_2=\rho_1\{1+\frac{M_0}{[-\frac{b}{2}+\sqrt{1+\frac{b^2}{4}+(\gamma-1)
(\frac{M_0^2}{2}-q)}]}\}.
\end{equation}

\subsection{Simulation results}
Considering the head-on collision of two gas sheets, the particles
with a positive $x$-coordinate are given an initial negative
velocity of Mach~5, and those with a negative $x$-coordinate are
given a Mach~5 velocity in the opposite direction.  The chosen
physical scales for length and time are $[l]=3.0 \times 10^{16}
\mathrm{m}$ and $[t]=3.0 \times 10^{13} \mathrm{s}$,
respectively, so the velocity unit is approximately
$1~\mathrm{km.s^{-1}}$. The gravity constant is set $G= 1$ for
which the calculated mass unit is $[m]=4.5 \times 10^{32}
\mathrm{kg}$. There is considered two equal one dimensional
molecular sheets with extension $x= 0.1 [l]$, which have initial
uniform density and temperature of $\sim 4.5\times 10^9
\mathrm{m^{-3}}$ and $\sim 10 \mathrm{K}$, respectively.

In adiabatic shock, with $M_0=5$, the post-shock density must be
$2.9$, which is obtained from analytic solution (\ref{e:den}) with
$Q=0$ and $\gamma=2$. On the other hand, in isothermal shock with
$M_0=5$, the post-shock density must be $26.9$ (i.e., from
Eq.~\ref{e:den} with $\gamma=1$). The simulation results of
adiabatic and isothermal shocks which use the MM97's AV, equation
(\ref{morris}), are shown in Fig.~1. The post-shock density has
oscillations for both cases of the adiabatic and isothermal
shocks. Since the thermal energy in adiabatic shock is restrained
in the system, the oscillations grow further in it as shown in
Fig.~1. Adding the linear AV term (\ref{linvisc}) can smooth the
oscillations, thus, we use the parameter $z_\alpha$ in the
$\alpha$ viscosity equation (\ref{alpha}). The best values of the
$z_\alpha$ for adiabatic and isothermal shocks are $2.0$ and
$3.5$, respectively. The value of $z_\alpha$ in adiabatic case is
small for the reason that the oscillations are greater than
isothermal case.

As shown in Fig.~1, the post-shock density of adiabatic case is
approximately equal to the analytical value, while the difference
in isothermal case is more distinguishable. This is because of
extremely increasing of the pressure in the compression region
that is greatly produced by the $\beta$-viscous term of AV in
equation (\ref{monag}). To remove this problem we propose using of
the parameter $z_\beta$ in the $\beta$ viscosity equation
(\ref{beta}). Choosing the values $0.5$ and $0.01$ for $z_\beta$
in adiabatic and isothermal case, respectively, can regulate the
post-shock density as shown in Fig.~2. These relative values are
because the pressure produced in the post-shock region of
isothermal case is artificially greater than the same pressure at
adiabatic case.

\section{Summary and conclusions}
In SPH method two forms of AV is necessary; the von
Neumann-Richtmeyer viscosity and linear viscous term. In this
work we revisited the relevant prescriptions for some AV, which
was proposed to consider these two forms of viscosity in SPH
methodology. As a good idea, MM97 introduced the time-varying
switch which fits more naturally with the particle formulation.
The shock formations in adiabatic and isothermal cases have been
performed to test the effect of different AVs.

The simulation results in which the MM97's AV was used, show that
the post-shock density has oscillations for both cases of the
adiabatic and isothermal shocks. The oscillations in adiabatic
shock grow further because the thermal energy in this case is
restrained in the system. On the other hand, the post-shock
density of adiabatic case is approximately equal to the
analytical value, while the difference in isothermal case is more
distinguishable.

To partly overcome some of these problems, we proposed a
modification of the AV. Based on the method of MM97, we used
time-variable $\alpha$ and $\beta$ in the standard AV. In this
approach, the AV is used only in supersonic regions where the gas
is not under compression, thus, the source term is restricted.
Using this source multiplied by two regulating parameters
($z_\alpha$ and $z_\beta$) can virtually eliminate some
problematic effects of using an artificial dissipation in SPH
capability of shock simulation. Computer experiments have been
performed to test the abilities of the proposal, and the results
show that the accuracy of these restrictions on AV is acceptable.
Although there are some satisfying results from shock formation
tests, the proposed AV should be communicated to the
astrophysical community, so that others can try it. There must be
other tests to complete the idea.


\clearpage
\begin{figure}
\epsscale{.70} \plotone{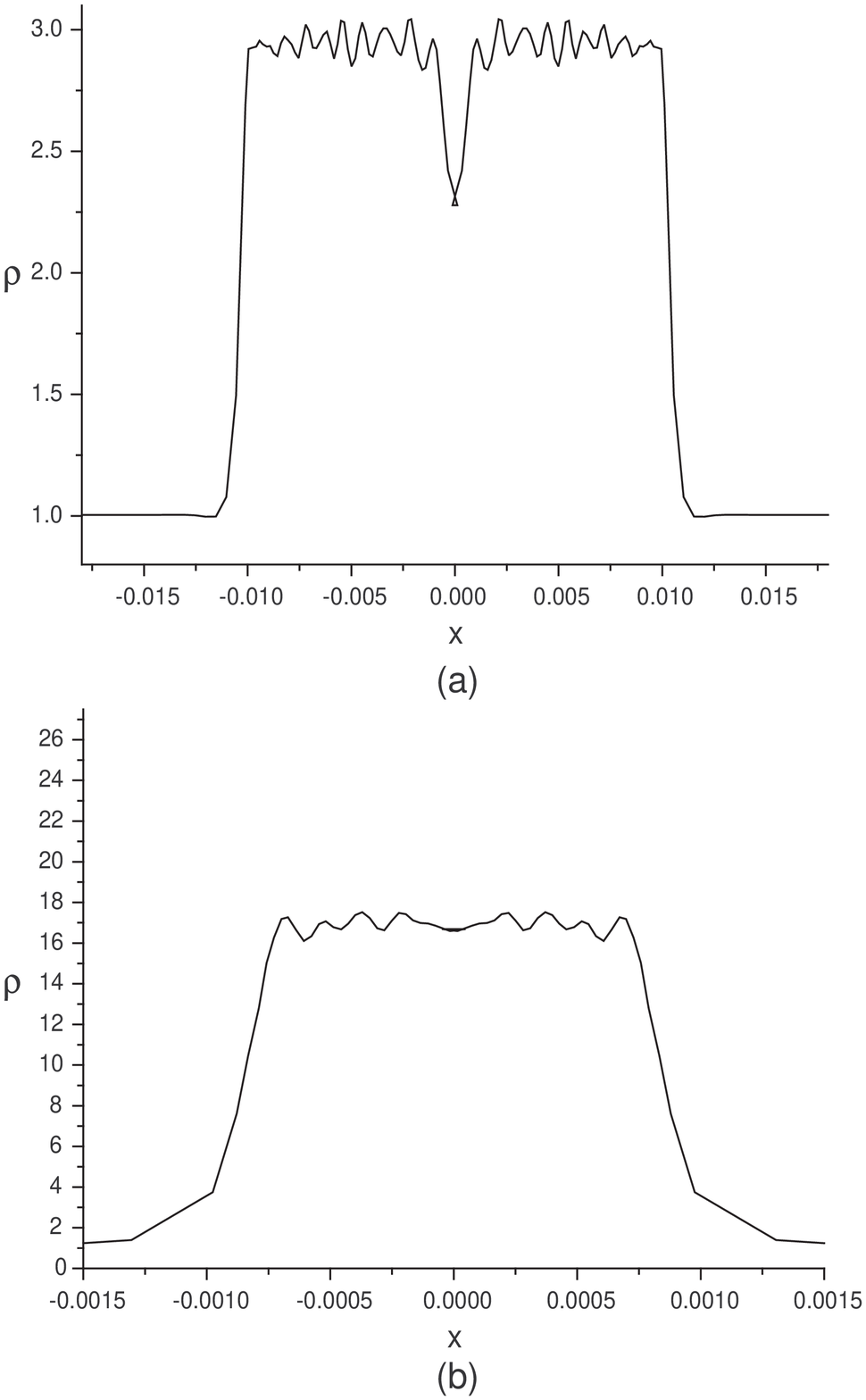} \caption{The density of (a)
adiabatic and (b) isothermal shocks in which the MM97's AV,
equation (\ref{morris}), was used.}
\end{figure}
\clearpage
\begin{figure}
\epsscale{.70} \plotone{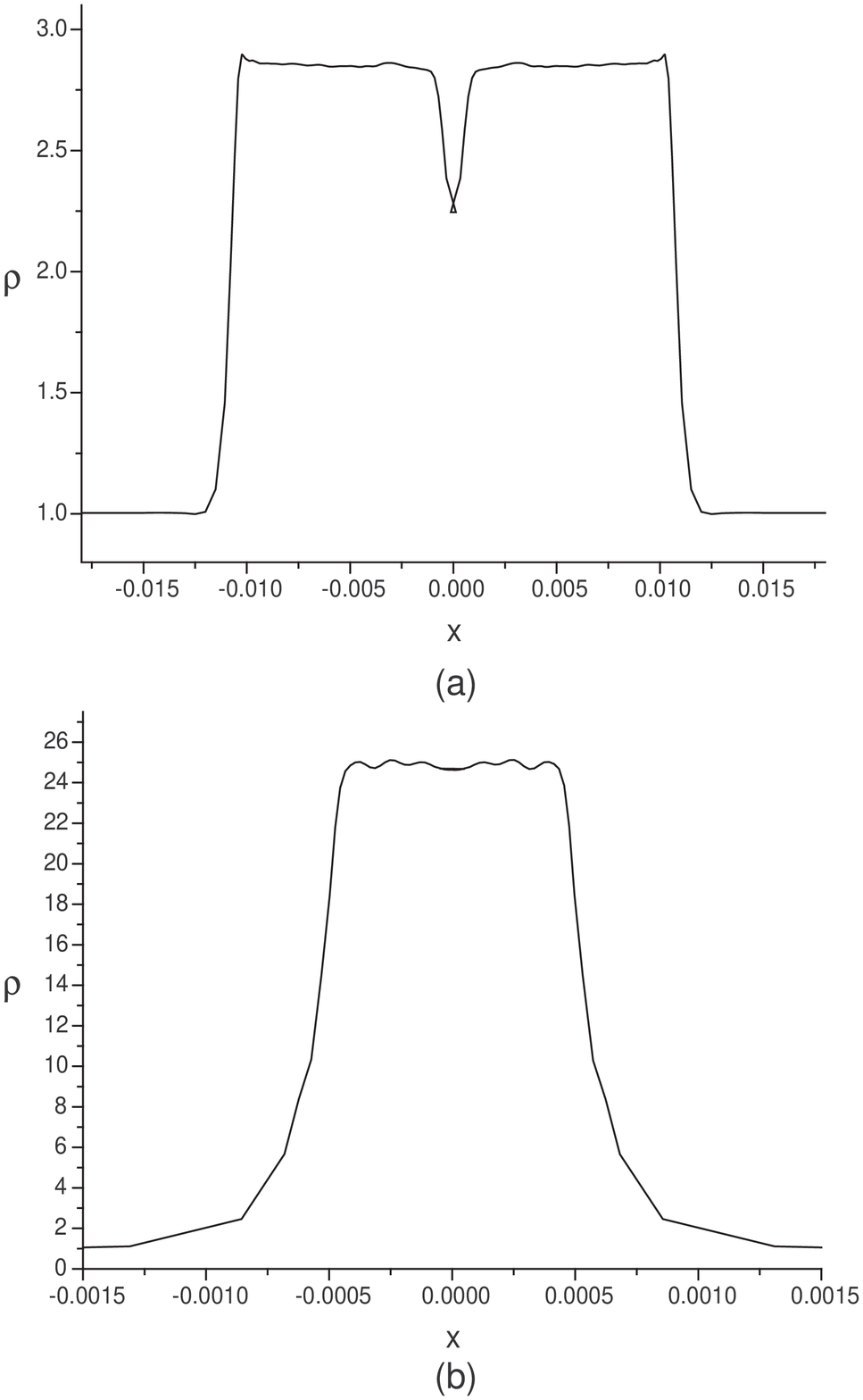} \caption{The density of (a)
adiabatic and (b) isothermal shocks in which the time variable
$\alpha$ and $\beta$ as outlined by equations (\ref{alpha}) and
(\ref{beta}), were used.}
\end{figure}

\end{document}